\documentclass[pra,showpacs,twocolumn,amsmath]{revtex4-1}
\usepackage{pifont}
\usepackage{graphicx,epsfig,subfigure,dsfont,amssymb,amsmath,amsthm,amsfonts,amsbsy,mathrsfs,amscd}

\usepackage[all]{xy}

\DeclareMathOperator{\tr}{tr}

\def\la{\langle}
\def\ra{\rangle}

\input amssym.def

\begin{document}

\title{Accessible Coherence and Coherence Distribution}

\smallskip
\author{Teng Ma$^1$ }
\author{Ming-Jing Zhao$^2$}
\author{Hai-Jun Zhang$^{3,4}$}
\author{Shao-Ming Fei$^{4,5}$}
\author{Gui-Lu Long$^1$}

\affiliation{$^1$State Key Laboratory of Low-Dimensional Quantum Physics and Department of Physics, Tsinghua University, Beijing 100084, China\\
$^2$School of Science,
Beijing Information Science and Technology University, Beijing, 100192, China\\
$^3$College of the Science, China University of Petroleum, Qingdao, 266580, China\\
$^4$School of Mathematical Sciences, Capital Normal
University, Beijing
100048, China\\
$^5$Max-Planck-Institute for Mathematics in the Sciences, 04103
Leipzig, Germany}


\begin{abstract}
The definition of accessible coherence is proposed. Through local measurement on the other subsystem and one way classical communication, a subsystem can  access more coherence than the coherence of its density matrix. Based on the local accessible coherence, the part that can not be locally accessed is also studied, which we call it remaining coherence. We study how the bipartite coherence is distributed by  partition  for both $l_1$ norm coherence and relative entropy coherence, and the expressions for local accessible coherence and remaining coherence are derived. we also study some examples to illustrate the distribution.
\end{abstract}

\maketitle

\section{Introduction}

Quantum coherence is one of the key features of the quantum world. It is the origin of many quantum phenomena such as the interference of light, the laser \cite{mandel}, superconductivity \cite{london}, and quantum thermodynamics \cite{oppenheim}. Recent researches show that quantum coherence also increases the efficiency of photosynthetic light-harvesting complexes \cite{yccheng, mohan}. Coherence is also the key ingredient in quantum computation and quantum information processing \cite{nielsen}. Recently, quantifying coherence as a resource in the context of quantum information science has been extensively investigated \cite{Baumgratz}. Many different coherence measures have been proposed and their properties have been studied \cite{napoli,jiajunma,yrzhang,shumingcheng,uttam,rana}. The relations between coherence and quantum correlations like quantum entanglement \cite{horodecki} and quantum discord \cite{discordo,discord} are also a research hot spot \cite{chitambar1,streltsov1,killoran,chitambar,streltsov,mateng,zhengjunxi}.

The quantum entanglement and quantum discord characterize the correlations between quantum systems. Basically these correlations are due to the superposition of quantum states, i.e., the quantum coherence.
Nevertheless, the quantum coherence exists in a single quantum system. In this sense, quantum coherence is more fundamental. As quantum correlations like entanglement satisfy trade-off relations among multipartite
systems \cite{monogamy}, it is also natural to ask how coherence is distributed among the subsystems.

In this paper, we study the distribution of coherence for bipartite systems. Different from the point of \cite{radhakri}, which regards the entanglement as the intrinsic coherence between subsystems, we study the distribution of coherence from the point of view of local accessible coherence. As an illustration, let us consider a bipartite state $1/2|+\ra \la + |\otimes|0\ra \la 0|+1/2|-\ra \la -|\otimes |1\ra \la 1|$ with $|+\ra=1/\sqrt{2}(|0\ra+|1\ra)$ and  $|-\ra=1/\sqrt{2}(|0\ra-|1\ra)$. This bipartite state has nonzero coherence under local computational basis. However, the reduced density matrices of the state are both identities: both subsystems have no coherence. The non-zero coherence of the whole system is due to the states $|+\ra$ and $|-\ra$ in the first system. However, the coherence is eliminated in the reduced state of the first system. We show that contributing to the coherence of the whole bipartite system are the coherence of the subsystems and the `accessible' coherence of the subsystems, together with the `remaining' coherence. We first propose the concept of accessible coherence, and analyze the property of accessible coherence. Then we study the property of the `remaining' coherence, the part that besides all the coherence and accessible coherence of the subsystems. At last, we illustrate how bipartite coherence is distributed by computing the accessible coherence and remaining coherence in detailed examples.

Throughout our paper, we take the reference basis to be the local computational basis and two important coherence measures, the relative entropy coherence  and the $l_1$ norm coherence \cite{Baumgratz} will be used.
The $l_1$ norm coherence $C^{l_1}$ of a quantum state $\rho$ is defined by the sum of the absolute value of the density matrix's off-diagonal elements,
\begin{equation}\label{cl1}
C^{l_1}(\rho)=\sum_{i\neq j} |\rho_{ij}|,
\end{equation}
where $|\rho_{ij}|$ is the absolute value of entry $\rho_{ij}$ of the density matrix $\rho$ under the reference basis.
The relative entropy coherence is defined by $C^r(\rho)=\min_{\sigma \in \mathcal{I}} S(\rho||\sigma)$, where $S(\rho||\sigma)$ is the relative entropy of states $\rho$ and $\sigma$, $\mathcal{I}$ denotes the set of incoherent states. $C^r(\rho)$ has a simple form,
\begin{equation}\label{cr}
C^r(\rho)=S(\rho_d)-S(\rho),
\end{equation}
where $S(\rho)=-\tr(\rho \log \rho)$ is the von Neumann entropy of $\rho$, and $\rho_d$ is the matrix of $\rho$ eliminating all the off-diagonal elements.

\section{Accessible coherence}

Consider a quantum system with state $\rho$, we define the accessible coherence of $\rho$ as the difference between the maximum   average coherence of its state ensemble and the coherence of the state,
\begin{equation}\label{accessiablecohmax}
\mathcal{C}^\mathcal{A}(\rho)=\max_{\{p_i, \rho_i\}} \sum_i p_i C(\rho_i)-C(\rho),
\end{equation}
where the maximization is taken over all state decompositions of $\rho=\sum_i p_i \rho_i$.
The accessible coherence is always positive since coherence measure is nonincreasing under mixing of quantum states \cite{Baumgratz}, i.e., $C(\rho)\leq \sum_i p_i C(\rho_i)$. Hence the coherence for a state is always smaller than or equal to the minimal value of the state's average coherence, and the accessible coherence quantifies the maximum  extra coherence one can gain when one knows the corresponding  ensemble of the state.


Due to the convexity of the coherence measure and the compact convexity of the density matrix set, the maximum in  (\ref{accessiablecohmax}) can actually be achieved by taking over all pure-state decompositions, i.e., $\max \sum_i p_i C(\rho_i)=\max \sum_i p'_i C(|\psi_i\ra)$, with $\rho=\sum_i p'_i|\psi_i\ra \la \psi_i|$. Then the accessible coherence of $\rho$ actually equals to its coherence of assistance \cite{chitambar1} minus its  coherence, and achieves its minimum, zero, when  the state $\rho$ is pure.
Note that the coherence of assistance is not a bonafide  measure  of quantum coherence for a single system \cite{streltsov-review}, neither is the accessible coherence then. Take a zero coherence state, $1/2 \mathbb{I}=1/2(|+\ra \la +|+|-\ra \la -|)$, with $|+\ra=1/\sqrt{2}(|0\ra+|1\ra)$ and  $|-\ra=1/\sqrt{2}(|0\ra-|1\ra)$, for example, it is easy to see that the state's accessible coherence and  the coherence of assistance are all nonzero, which violates the conditions for a bonafide measure of coherence \cite{Baumgratz}. However, the accessible coherence reveals the difference between a state's coherence of assistance and the quantum coherence itself.

We have defined the accessible coherence in (\ref{accessiablecohmax}) with a maximization over all ensembles of a state decompositions. For a specific ensemble of $\rho=\sum_i p_i \rho_i$, we denote
\begin{equation}\label{accessiablecoh}
C^\mathcal{A}(\rho)=\sum_i p_i C(\rho_i)-C(\rho),
\end{equation}
the accessible coherence for the specific ensemble.
Under the relative entropy measure, the accessible coherence is tightly connected to the Holevo quantity for the accessible information \cite{nielsen}.

Suppose there is a sender, Alice, who prepares states $\rho_i$ with probabilities $p_i$. Through a channel, she sends these states to a receiver, Bob, who performs a POVM measurement to distinguish these states to get as much information as he can. If there is no noise in the channel, the mutual information between Alice and Bob is bounded by the Holevo quantity:
\begin{equation}\label{holevo}
\chi(\rho)=S(\rho)-\sum_i p_i S(\rho_i),
\end{equation}
where $\rho=\sum_i p_i \rho_i$.  If Alice sends the states through a decoherence channel $\Pi(\cdot)=\sum_i|i\ra \la i|(\cdot)|i\ra \la i|$, then the Holevo quantity decreases to
\begin{equation}\label{holevo1}
\chi'(\rho)=S(\Pi(\rho))-\sum_i p_i S(\Pi(\rho_i)),
\end{equation}
where $\Pi(\rho)=\sum_i p_i \Pi(\rho_i)$. Using (\ref{cr}), (\ref{accessiablecoh}), and $\Pi(\rho)=\rho_d$, we get
\begin{equation}
C^\mathcal{A}(\rho)=\chi(\rho)-\chi'(\rho),
\end{equation}
from (\ref{holevo}) and (\ref{holevo1}).
Hence we see that the accessible coherence with respect to the information source's  ensemble equals to the decrease of Holevo quantity due to the channel's decoherence. This relation holds also for $\mathcal{C}^\mathcal{A}(\rho)$.

The accessible coherence plays roles in the distribution of coherence.
As the accessible coherence with respect to a specific ensemble comes from the information of the ensemble, for a bipartite system AB, one way to gain the ensemble of system A is to measure  system B and communicate each measurement outcome to A \cite{yuchangshui,wootters}. In this way, the system A can gain an averaged coherence which is lager than the coherence of its density matrix, denoting as $C_A$, and the extra part is the accessible coherence of system A, denoting as $C^\mathcal{A}_A$. The similar analysis also holds for system B. In the following we consider the local measurement under the reference basis.

From the analysis above, the coherence contained in system A (B) are given by $C_A$ and $C^\mathcal{A}_A$ ($C_B$ and $C^\mathcal{A}_B$). The total coherence of the whole system AB is from
$C_A, C^\mathcal{A}_A, C_B, C^\mathcal{A}_B$ and some `remaining' coherence, denoting as $C^\mathcal{T}$, between A and B. $C^\mathcal{T}$ represents the part that can not be locally accessed  under local reference basis measurement and one way classical communication. Hence from the point of view of local accessible coherence, we have the partition for the bipartite coherence $C_{AB}$ of $\rho_{AB}$,
\begin{equation}\label{partition1}
C_{AB}=C_A+C^\mathcal{A}_A+C_B+C^\mathcal{A}_B+C^\mathcal{T}.
\end{equation}
Later, we will show that the remaining coherence $C^\mathcal{T}$ is always non-negative for relative entropy coherence and $l_1$ norm coherence measures.
Basically $C^\mathcal{T}$ is due to correlations between systems A and B. In the following we investigate and calculate the quantities $C_A$, $C^\mathcal{A}_A$, $C_B$, $C^\mathcal{A}_B$ and $C^\mathcal{T}$ in detail.

\section{Coherence distribution with relative entropy measure}

The first thing we notice is the connection between coherence and the measurement dependent quantum discord $D$ \cite{discord} with the measurement basis being the local reference  basis. Here $D$ is distinguished from the original discord \cite{discordo,discord}. The unilateral quantum discord $D_{AB}^\leftarrow$ with respect to the local projective measurement $\Pi(\cdot)=\sum_i|i\ra \la i|(\cdot)|i\ra \la i|$ on system B is defined to be the difference of mutual information $I_{AB}$ and the mutual information $I_{A \widetilde{B}}$ after the measurement on B \cite{discord},
\begin{equation}\label{disc}
\begin{aligned}
D_{AB}^\leftarrow=& I_{AB}-I_{A \widetilde{B}}\\
=&S_A+S_B - S_{AB}-S_A- S_{\widetilde{B}} + S_{A\widetilde{B}}\\
=&S_B - S_{AB}- S_{\widetilde{B}} + S_{A\widetilde{B}},
\end{aligned}
\end{equation}
where `$^\leftarrow$' means that the measurement takes on system B, $S_X$ denotes the von Neumann entropy of the system $X$ and $ \widetilde{X} $  denotes the measurement $\Pi$ on system $X$.
Similarly, we have unilateral discord $D_{AB}^\rightarrow=I_{AB}-I_{\widetilde{A}B}$ and bilateral discord $D_{AB}^\leftrightarrow=I_{AB}-I_{\widetilde{A} \widetilde{B}}$, where `$^\rightarrow$' and `$^\leftrightarrow$' denote the measurements on system A and both systems AB  respectively.

The unilateral and bilateral (bipartite)  coherence \cite{jiajunma} is defined as
\begin{equation}\nonumber\begin{aligned}
&C^{r\rightarrow}=S_{\widetilde{A}B}-S_{AB},\\
&C^{r\leftarrow}=S_{A\widetilde{B}}-S_{AB},\\
&C^{r\leftrightarrow}=S_{\widetilde{A}\widetilde{B}}-S_{AB},
\end{aligned}\end{equation}
which correspond to the measurements on system A, system B, and both systems AB  respectively.
From (\ref{cr}), (\ref{disc}) and $\rho_{B_d}=\Pi(\rho_B)$, we have unilateral coherence with respect to the measurement on B,
$C^{r\leftarrow}=C^r_B+D_{AB}^\leftarrow$.
 If $C^{r\leftarrow}=0$, which means that the state is a quantum-incoherent state \cite{streltsov,bromley}, i.e., $\rho_{AB}=\sum_i p_i \rho^A_i\otimes|i\ra \la i|$, the  coherence of system B is zero, and there is no right side quantum correlation $D_{AB}^\leftarrow$ between A and B. Similarly, we have unilateral coherence with respect to the measurement on A,
$C^{r\rightarrow}=C^r_A+D_{AB}^\rightarrow$.
Following the above steps we also have the coherence with respect to bipartite local measurements,
\begin{equation}\label{cd}
 C^{r\leftrightarrow}=C^r_A+C^r_B+D_{AB}^\leftrightarrow.
 \end{equation}
From equations (\ref{partition1}) and (\ref{cd}), we see that the local accessible coherence and the remaining coherence are related to the quantum discord,
\begin{equation}\label{dcon}
D_{AB}^\leftrightarrow=C^{\mathcal{T}^r}+C^{\mathcal{A}^r}_A+C^{\mathcal{A}^r}_B.
\end{equation}

Next, we find out the expressions for $C^{\mathcal{T}^r}$, $C^{\mathcal{A}^r}_A$, $C^{\mathcal{A}^r}_B$, and their relations with the quantum correlations $D$ and coherence $C^r$.
Note that $I_{A\widetilde{B}}$ in (\ref{disc}) is actually the classical correlation  \cite{discordo,discord} (the Holevo quantity for system A), which can be reexpressed as
\begin{equation}\label{jleft}
I_{A\widetilde{B}}=S_A-\sum_i p_i S(\rho_{A_i}),
\end{equation}
where the state $\rho_{A_i}$ with probability $p_i$ corresponds to system B's measurement outcome $i$.
If one applies  measurement $\Pi$ on system A too, one gets the bilateral classical correlation,
\begin{equation}\label{jleftright}
I_{\widetilde{A}\widetilde{B}}=S_{\widetilde{A}}-\sum_i p_i S(\widetilde{\rho}_{A_i}),
\end{equation}
which corresponds to the bilateral discord $D_{AB}^\leftrightarrow$.
Equations (\ref{jleft}) and (\ref{jleftright}) give rise to
\begin{equation}\label{cping}
\sum_i p_i C^r(\rho_{A_i})-C^r_A=D_{AB}^\leftrightarrow-D_{AB}^\leftarrow.
\end{equation}
 From (\ref{accessiablecoh}), (\ref{dcon}) and (\ref{cping}) we have
$D_{AB}^\leftarrow=C^{\mathcal{T}^r}+C^{\mathcal{A}^r}_B$,
and similarly
$D_{AB}^\rightarrow=C^{\mathcal{T}^r}+C^{\mathcal{A}^r}_A$.
From the definition  (\ref{disc}) and the relation (\ref{cping}), we have the expression of the local accessible coherence for system A,
\begin{equation}\begin{aligned}
C^{\mathcal{A}^r}_A&=D_{AB}^\leftrightarrow-D_{AB}^\leftarrow\\
&=S_{\widetilde{A}\widetilde{B}}-S_{A\widetilde{B}}+S_A-S_{\widetilde{A}}.
\end{aligned}
\end{equation}
Similarly, for system B we also have
\begin{equation}\begin{aligned}
C^{\mathcal{A}^r}_B&=D_{AB}^\leftrightarrow-D_{AB}^\rightarrow\\
&=S_{\widetilde{A}\widetilde{B}}-S_{\widetilde{A}B}+S_B-S_{\widetilde{B}}.
\end{aligned}
\end{equation}

Next we compute the remaining coherence $C^{\mathcal{T}^r}$ in terms of the relative entropy measure of coherence and discuss its physical implications.
From the definition (\ref{disc}), the bilateral discord  can be split into two parts,
$D^\leftrightarrow_{AB}
=D^\leftarrow_{AB}+D^\rightarrow_{A\widetilde{B}}=D^\rightarrow_{AB}+D^\leftarrow_{\widetilde{A}B}$. Therefore we have the following expression for the remaining coherence,
\begin{equation}\begin{aligned}\label{ci1}
C^{\mathcal{T}^r}&=D_{AB}^\rightarrow-D_{A\widetilde{B}}^\rightarrow=D_{AB}^\leftarrow-D_{\widetilde{A}B}^\leftarrow\\
&=S_{A\widetilde{B}}+S_{\widetilde{A}B}-S_{AB}-S_{\widetilde{A}\widetilde{B}}.
\end{aligned}
\end{equation}
Since the quantum discord quantifies the change of mutual information induced by measurement, equation (\ref{ci1}) indicates that the remaining coherence quantifies a kind of decrease of quantum discord:
 The quantum discord, i.e., the amount of  mutual information revealed by a measurement on system A,
$D_{AB}^\leftarrow$, decreases if another measurement on system B has already revealed some mutual information.

The remaining coherence, $C^{\mathcal{T}^r}$, is always non-negative. From (\ref{ci1}) we have
\begin{equation}\begin{aligned}\label{cipositive}
C^{\mathcal{T}^r}&=(S_{A\widetilde{B}}-S_{AB})-(S_{\widetilde{A}\widetilde{B}}-S_{\widetilde{A}B})\\
&=S(\rho_{AB}||\rho_{A\widetilde{B}})-S(\rho_{\widetilde{A}B}||\rho_{\widetilde{A}\widetilde{B}})\geq0,
\end{aligned}
\end{equation}
where the second equality is due to the fact that for projective operator $\Pi$ and matrices $A$ and $B$, $\tr [(\Pi A) B]=\tr [A \Pi (B)]=\tr [\Pi(A) \Pi(B)]$, and the inequality is due to that the relative entropy is contractive under any completely positive trace preserving map \cite{lindblad,Wehrl}.

Moreover, for any incoherent-incoherent, quantum-incoherent and incoherent-quantum state \cite{streltsov,bromley}, $C^{\mathcal{T}^r}=0$. An open question is whether the incoherent-incoherent, quantum-incoherent and incoherent-quantum states are the necessary and sufficient conditions such that the equality in (\ref{cipositive}) holds. Note that the remaining coherence is different from the intrinsic coherence in \cite{radhakri}, which actually is the entanglement. It can be shown that only for some states the remaining coherence is equal to the entanglement (see state (\ref{schmidtst}) in  examples).

\section{Coherence distribution with $l_1$ norm measure}

Let us first analyze the coherence of a bipartite state $\rho_{AB}$ in terms of its entries
under local reference basis,
\begin{equation}\label{rhoab}
\rho_{AB}=\sum_{ikjl}\rho_{ik,jl}|i\ra \la j|\otimes |k\ra \la l|.
\end{equation}
The entries $\rho_{ik,jk (i\neq j)}$ contribute the  coherence for the system A only. However, the coherence induced by $\rho_{ik,jk (i\neq j)}$ may be canceled when we only consider the marginal state $\rho^A=\tr_B \rho_{AB}$, since there may exist some entries such that $\rho_{ik',jk'(i\neq j)}=-\rho_{ik,jk(i\neq j)}$. This indicates that the coherence induced by the entries  $\rho_{ik,jk(i\neq j)}$ can be divided into two parts: The first part is not canceled and still lives in the density matrix $\rho^A$ after tracing over the subsystem B, and the second part is canceled in $\rho_A$. As we will  show, the first part actually is the $l_1$ norm coherence of $\rho_A$,  denoting as $C_A^{l_1}$, and the second part actually is  the local accessible coherence of system A, denoting as $C^{\mathcal{A}^{l_1}}_A$. The same analysis also holds for entries  $\rho_{ik,il(k\neq l)}$, which correspond to the coherence partition for the system B. Hence the above analysis of the entries of density matrix also shows that, for a subsystem, there are some accessible coherence besides the coherence of the subsystem's density matrix.

Next we find out the expressions for  $C^{\mathcal{A}^{l_1}}_A$, $C^{\mathcal{A}^{l_1}}_B$, and $C^{\mathcal{T}^{l_1}}$.
For the state (\ref{rhoab}), we have the bipartite coherence for $l_1$ norm measure,
\begin{equation}\label{rhocl1}
C^{l_1}_{AB}=\sum_{i\neq j,k\neq l} |\rho_{ik,jl}|+\sum_{i\neq j,k} |\rho_{ik,jk}|+\sum_{i,k\neq l} |\rho_{ik,il}|.
\end{equation}

Under local measurement $\Pi$ on system B, the state of system A is given by $\rho_{A_k}=\sum_{ij}\rho_{ik,jk}/p_k |i\ra \la j|$, with probability  $p_k=\sum_{i}\rho_{ik,ik}$.
For the middle term of the right hand side of (\ref{rhocl1}), from (\ref{accessiablecoh}) it is easy to check
\begin{equation}\label{cl1ping}
\sum_{i\neq j,k} |\rho_{ik,jk}|=\sum_k p_k C^{l_1}(\rho_{A_k})=C^{l_1}_A+C^{\mathcal{A}^{l_1}}_A.
\end{equation}
The above equation means that, for the $l_1$ norm coherence, the total coherence induced only by system A's basis, $\sum_{i\neq j,k} |\rho_{ik,jk}|$, equals to the average coherence of A, which is the summation of coherence and accessible coherence of system A. This result is in consist with our former  analysis according to the entries of density matrix. From (\ref{cl1ping}) we get the expression of accessible coherence for system A,
\begin{equation}\label{cl1a}
C^{\mathcal{A}^{l_1}}_A=\sum_{i\neq j}(\sum_k |\rho_{ik,jk}|-|\sum_k \rho_{ik,jk}|).
\end{equation}
Similarly, we have
that the average coherence of B equals to the summation of coherence and accessible coherence of system B. And the expression of accessible coherence for system B is given by
\begin{equation}\label{cl1b}
C^{\mathcal{A}^{l_1}}_B=\sum_{k\neq l}(\sum_i |\rho_{ik,il}|-|\sum_i \rho_{ik,il}|).
\end{equation}

The remaining coherence can be obtained from (\ref{partition1}), (\ref{rhocl1}), (\ref{cl1a}), and (\ref{cl1b}),
\begin{equation}
C^{\mathcal{T}^{l_1}}=\sum_{i\neq j,k\neq l} |\rho_{ik,jl}|.
\end{equation}
Obviously, $C^{\mathcal{T}^{l_1}}$ is always non-negative. Since the entries $ \rho_{ik,jl}$ with $i\neq j$ and $k\neq l$ represent the correlation between the two subsystems, the remaining coherence $C^{\mathcal{T}^{l_1}}$ quantifies all the correlation induced by  these entries. However, $C^{\mathcal{T}^{l_1}}$ is different from the entanglement in general, although it equals to entanglement for some states (see examples).

Note that the zero points of the remaining coherence for the relative entropy measure and $l_1$ norm measure may be different. For states which do not contain entries $\rho_{ik,jl}$ with $i\neq j$ and $k\neq l$, $C^{\mathcal{T}^{l_1}}=0$, but possibly $C^{\mathcal{T}^r}\neq 0$.

\section{Examples}
\begin{figure}
\centering
\subfigure[~Relative entropy coherence]{
\label{fig:subfig:a}
\includegraphics[width=1.5in]{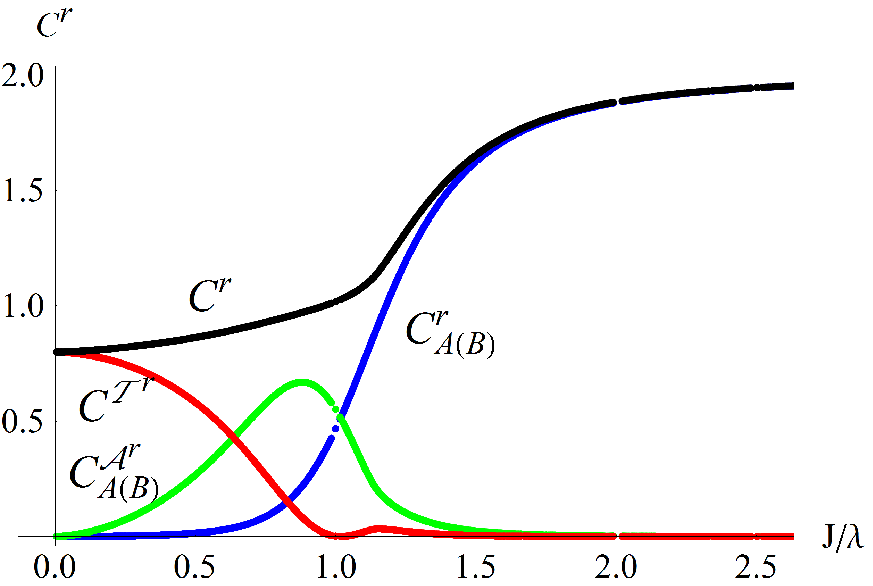}}
\hspace{0.1in}
\subfigure[~$l_1$ norm coherence]{
\label{fig:subfig:b}
\includegraphics[width=1.5in]{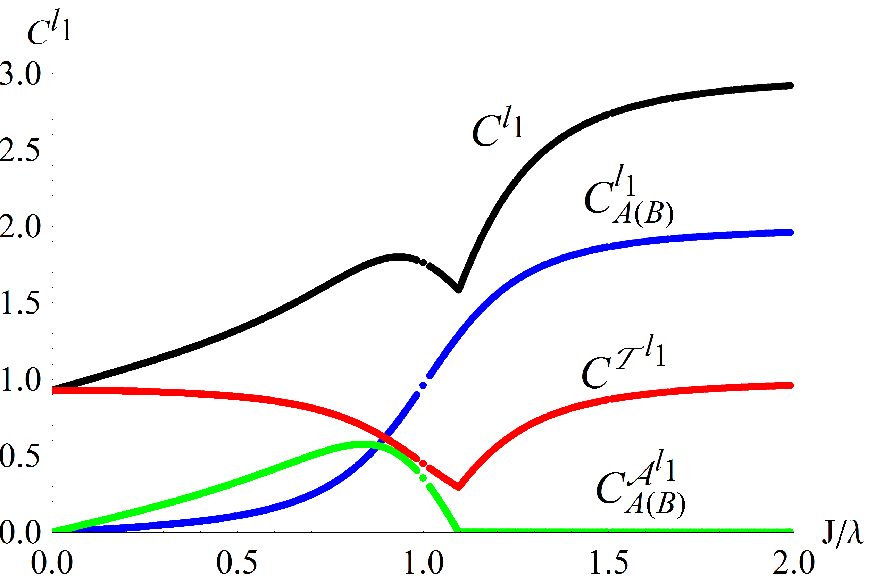}}
\caption{The distribution of coherence for the ground  state of (\ref{ising})
 with  relative entropy measure (a) and the $l_1$ norm measure (b).
 At $J/\lambda=0$, for both relative entropy coherence and $l_1$ norm coherence, the local coherence ($C^{r}_{A(B)}$ and $C^{l_1}_{A(B)}$, blue dots) and local accessible coherence ($C^{\mathcal{A}^{r}}_{A(B)}$ and $C^{\mathcal{A}^{l_1}}_{A(B)}$, green dots) are all zero,
and  the bipartite coherence ($C^{r}$ and $C^{l_1}$, black dots) is just the remaining coherence ($C^{\mathcal{T}^{r}}$ and $C^{\mathcal{T}^{l_1}}$, red dots). At $J/\lambda\rightarrow\infty$,
for the relative entropy coherence, the local accessible coherence and the remaining coherence approach to zero. And all the bipartite coherence is from local coherence. While for the $l_1$ norm coherence
the remaining coherence is nonzero, and the bipartite coherence is distributed into both local
coherence and the remaining coherence.}
\label{fig:subfig} 
\end{figure}

We now consider some examples for the partition of coherence in bipartite systems, and show how the local accessible coherence and the remaining coherence are distributed for both relative entropy coherence and $l_1$ norm coherence. First let us consider the Schmidt correlated state \cite{rains},
\begin{equation}\label{schmidtst}
\rho^s_{AB}=\sum_{ij}\rho_{ii,jj}|i\ra \la j|\otimes |i\ra \la j|.
\end{equation}
The marginal states of  (\ref{schmidtst}) are all diagonal, thus the local coherence $C^r_A=C^r_B=C^{l_1}_A=C^{l_1}_B=0$. Since the state becomes diagonal  under local measurement $\Pi$ on systems A or B, its local accessible coherences are also zero, $C^{\mathcal{A}^r}_A=C^{\mathcal{A}^r}_B=C^{\mathcal{A}^{l_1}}_A=C^{\mathcal{A}^{l_1}}_B=0$.
 The remaining coherence $C^{\mathcal{T}^{l_1}}(\rho^s_{AB})=\sum_{i\neq j}|\rho_{ii,jj}|$ and $C^{\mathcal{T}^r}(\rho^s_{AB})=S(\rho^s_{\widetilde{A}\widetilde{B}})-S(\rho^s_{AB})$, which are actually the entanglement under negativity measure  and  relative entropy measure, respectively \cite{zhao}. Hence for the state (\ref{schmidtst}), the total coherence is contributed by the remaining coherence, here the entanglement between A and B.

For another example we consider the ground state of $N=2$ Ising model described by the Hamiltonian \cite{radhakri},
\begin{equation}\label{ising}
H=\lambda \sigma^x_1\sigma^x_1+J(\sigma^x_1+\sigma^x_2)+\epsilon \lambda (\sigma^z_1+\sigma^z_2),
\end{equation}
where $J,\lambda$ are the coupling parameters and $\epsilon$ is the symmetry-breaking term.
The ground state approaches to the Bell state $|00\ra+|11\ra/\sqrt{2}$ for $J\ll \lambda$ and the product state $(|0\ra+|1\ra)(|0\ra+|1\ra)/2$ for $J\gg \lambda$.
The distribution for both retaliative entropy coherence and  $l_1$ norm coherence are shown in Fig. 1.

\section{Conclusions}

We have proposed the concept of accessible coherence for a single quantum system and studied the properties of accessible coherence. The accessible coherence with respect to any given specific ensemble of a state has been shown to be connected to the decrease of the Holevo quantity due to the channel's decoherence.  The accessible coherence also shows its role in the coherence distribution in bipartite systems. The local accessible coherence of system A can be gained by measuring the system B and communicating the measurement results to A. We have studied the local accessible coherence and the unaccessible remaining coherence, for both relative entropy coherence and $l_1$  norm coherence. Their explicit expressions have been derived analytically. Under the local reference basis measurement, for both relative entropy coherence and $l_1$ norm coherence, we found that the local average coherence over all the measurement outcomes is the summation of local accessible coherence and coherence of the local reduced density matrix. We have shown that the remaining coherence in terms of relative entropy coherence quantifies the decrease of the basis dependent discord due to a local decoherence measurement. While the remaining coherence with $l_1$ norm measure equals to the sum of absolute values of the density matrix's entries with different subscripts for both parties. We finally studied how the bipartite coherence are distributed into local coherence, local accessible coherence and remaining coherence through some examples. The results give us an informational picture of coherence distributions in bipartite systems.

\smallskip
\noindent{\bf Acknowledgments}\, \,We thank Zhen Wang and Bao-Zhi Sun for useful discussions. This work is supported by the NSFC under numbers 11401032, 11675113, 11175094 and 91221205, and the National Basic Research Program of China (2015CB921002).


\begin{thebibliography}{99}
\bibitem{mandel}   L. Mandel and E. Wolf, Optical Coherence and Quantum
Optics (Cambridge University Press, Cambridge, England,
1995).

\bibitem{london} F. London and H. London, Proc. R. Soc. A \textbf{149}, 71
(1935).

\bibitem{oppenheim} P. \ifmmode \acute{C}\else \'{C}\fi{}wikli\ifmmode \acute{n}\else \'{n}\fi{}ski, M. Studzi\ifmmode \acute{n}\else \'{n}\fi{}ski, M. Horodecki,  and J. Oppenheim, Phys. Rev. Lett. \textbf{115}, 210403 (2015).


\bibitem{yccheng} Y. C. Cheng and G. R. Fleming, Annu. Rev. Phys. Chem. \textbf{60}, 241 (2009).
\bibitem{mohan} M. Sarovar, A. Ishizaki, G. R. Fleming, and K. B. Whaley, Nat. Phys. \textbf{6}, 462 (2010).
\bibitem{nielsen}M. A. Nielsen and L. Chuang, Quantum Computation
and Quantum Information (Cambridge University Press,
Cambridge, England, 2000).

\bibitem{Baumgratz} T. Baumgratz, M. Cramer, and M. B. Plenio, Phys. Rev. Lett. \textbf{113}, 140401 (2014).

\bibitem{napoli} C. Napoli, T. R. Bromley, M. Cianciaruso, M. Piani, N. Johnston, and G. Adesso, Phys. Rev. Lett. \textbf{116}, 150502 (2016).

\bibitem{jiajunma}J. J. Ma, B. Yadin, D. Girolami, V. Vedral, and M. Gu, Phys. Rev. Lett. \textbf{116}, 160407 (2016).

\bibitem{yrzhang} Y. R. Zhang, L. H. Shao, Y. Li, and H. Fan, Phys. Rev. A \textbf{93}, 012334 (2016).

\bibitem{shumingcheng} S.  Cheng and Michael J. W. Hall, Phys. Rev. A \textbf{92}, 042101 (2015).


\bibitem{uttam} U. Singh, L. Zhang, and A. K. Pati, Phys. Rev. A \textbf{93}, 032125 (2016).

\bibitem{rana} S. Rana, P. Parashar, and M. Lewenstein, Phys. Rev. A \textbf{93}, 012110 (2016).

\bibitem{horodecki}R. Horodecki, P. Horodecki, M. Horodecki, and K. Horodecki,
Rev. Mod. Phys. \textbf{81}, 865 (2009).

\bibitem{discordo}H. Ollivier and W. H. Zurek, Phys. Rev. Lett. \textbf{88},  017901 (2001).

\bibitem{discord} K. Modi, A. Brodutch, H. Cable, T. Paterek, and V. Vedral, Rev. Mod. Phys. \textbf{84}, 1655 (2012).

 \bibitem{chitambar1} E. Chitambar, A. Streltsov, S. Rana, M. N. Bera, G. Adesso, and M. Lewenstein, Phys. Rev. Lett. \textbf{116}, 070402 (2016).

\bibitem{streltsov1} A. Streltsov, E. Chitambar, S. Rana, M. N. Bera, A. Winter, and M. Lewenstein, Phys. Rev. Lett. \textbf{116}, 240405 (2016).

\bibitem{killoran} N. Killoran, F. E. S. Steinhoff, and M. B. Plenio, Phys. Rev. Lett. \textbf{116}, 080402 (2016).

\bibitem{chitambar} E. Chitambar and M. H. Hsieh, Phys. Rev. Lett. \textbf{117}, 020402 (2016).

\bibitem{streltsov}A. Streltsov, U. Singh, H. S. Dhar, M. N. Bera, and G. Adesso,
Phys. Rev. Lett. \textbf{115}, 020403 (2015).

\bibitem{mateng}T. Ma, M. J. Zhao, S. M. Fei, and G. L. Long, Phys. Rev. A \textbf{94}, 042312 (2016).

\bibitem{zhengjunxi} Z. J. Xi, Y. M. Li,  and H. Fan,  Sci. Rep. \textbf{5}, 10922 (2015).



\bibitem{monogamy}
V. Coffman, J. Kundu, W. K. Wootters, Phys. Rev. A \textbf{61}, 052306 (2000).\\
T. J. Osborne, and F. Verstraete, Phys. Rev. Lett. 96, 220503 (2006).\\
X. N. Zhu, and S. M. Fei, Phys. Rev. A \textbf{90}, 024304 (2014).\\
X. N. Zhu, and S. M. Fei, Phys. Rev. A \textbf{92}, 062345 (2015).

\bibitem{radhakri} C. Radhakrishnan, M. Parthasarathy, S. Jambulingam, and T. Byrnes, Phys. Rev. Lett. \textbf{116}, 150504 (2016).
\bibitem{streltsov-review} A. Streltsov, G. Adesso, and M. B. Plenio, Colloquium: Quantum Coherence as a Resource, arXiv:1609.02439v1.

\bibitem{yuchangshui} C. S. Yu and H. S. Song, Phys. Rev. A  \textbf{80}, 022324 (2009).
\bibitem{wootters} L. P. Hughston, R. Josza, and W. K. Wootters, Phys. Lett. A \textbf{183}, 14 (1993).
\bibitem{bromley}T. R. Bromley, M. Cianciaruso, and G. Adesso,
Phys. Rev. Lett. \textbf{114}, 210401  (2015).
\bibitem{lindblad} G. Lindblad, Commun. Math. Phys. \textbf{40}, 147 (1975).
\bibitem{Wehrl} A. Wehrl, Rev. Mod. Phys. \textbf{50}, 221 (1978).

\bibitem{rains} E. M. Rains, Phys. Rev. A \textbf{60}, 179 (1999).

\bibitem{zhao} M. J. Zhao, S. M. Fei, and Z. X. Wang, Phys. Lett.  A \textbf{372}, 2552 (2008).



\end{thebibliography}
\end{document}